\documentclass[lettersize,journal]{IEEEtran}

\usepackage{graphicx}
\usepackage{epstopdf}
\usepackage{float}
\usepackage{array}
\usepackage{amsmath}
\usepackage{amssymb}
\usepackage{amsthm}
\usepackage{eqparbox}
\usepackage{fixltx2e}
\usepackage{makecell}
\usepackage{mathrsfs}
\usepackage{amsmath}
\usepackage{multicol}

\usepackage{url}
\usepackage{cite}
\usepackage[justification=centering]{caption}
\usepackage{bm}
\usepackage{threeparttable}
\usepackage{subeqnarray}
\usepackage{cases}
\usepackage{color}
\usepackage{stfloats}
\usepackage[section]{placeins}
\usepackage{setspace}
\usepackage{algorithm, algorithmic}
\allowdisplaybreaks[4]

\ifCLASSOPTIONcompsoc
\usepackage[caption=false,font=normalsize,labelfont=sf,textfont=sf]{subfig}
\else
\usepackage[caption=false,font=footnotesize]{subfig}
\fi

\allowdisplaybreaks[4]

\newcommand{\diag}{\mathop{\mathrm{diag}}}



\begin{document}
	\captionsetup{font={small}}
	\title{Joint Precoder and Reflector Design \\ for RIS-assisted Multi-user OAM \\ Communication Systems }
	\author{Xiaoyan Ma,~\IEEEmembership{Member, IEEE,}
		Yufei Zhao,~\IEEEmembership{Member, IEEE,} 
		Haixia Zhang,~\\ \IEEEmembership{{Senior Member}, IEEE,}
	   Yong Liang Guan,~\IEEEmembership{Senior Member, IEEE,}		
	   \\ and Chau Yuen,~\IEEEmembership{Fellow, IEEE}

	
	}
	\maketitle
	\IEEEpeerreviewmaketitle
	\begin{abstract}
	Orbital angular momentum (OAM) can enhance the spectral efficiency by multipling a set of orthogonal modes on the same frequency channel. Utilizing the spatial orthogonality of different OAM modes, independent data streams can be simultaneously transmitted without interference, which enables OAM as a new form of multiple access technique for future wireless communications. Existing researches on OAM mainly focus on point-to-point transmissions in the line-of-sight (LoS) scenarios, since the
	perfect alignments between transmitters and receivers are strictly required to maintain the orthogonality between different OAM modes. However, in multi-user OAM communications, the perfect alignments between the transmitter and all the receivers are impossible.
	The phase turbulence, caused by misaligned transmitters and receivers, will lead to serious inter-mode interference, thereby the signals cannot be correctly detected at the receiver. To achieve non-line-of-sight (NLoS) OAM transmissions in misaligned scenarios, we investigate the joint precoder and reflector design for reconfigurable intelligent surface (RIS)-assisted multi-user OAM communication systems. Specifically,
	we propose a three-layer design at the transmitter side, which includes inter-user OAM mode interference cancellation, inter-mode self-interference elimination and the power allocation among different users. By analyzing the characteristics of the overall channels, we are able to give the specific expressions of the precoder designs, which significantly reduce the optimization complexity of the considered system. Furthermore, with the proposed three-layer precoder design, we successfully transfer the considered multi-user OAM communication system into a virtual MIMO system with distributed receiving antennas, and leveraging RIS to further enhance the sum rate performance with closed-from solutions. To verify the superiority of the proposed multi-user OAM transmission system, we compare it with traditional MIMO transmission schemes, numerical results have shown that our proposed design can achieve better sum rate performance due to the well-designed orthogonality among different receivers and OAM modes.
	\end{abstract}	
	

	\section{Introduction}
Orbital angular momentum (OAM) communications have gained lots of attentions due to the ability of increasing the spectral efficiency for future wireless communication networks. By utilizing the spatial orthogonality of helical phase, OAM can carry independent data streams on differnent OAM modes to achieve interference-free transmissions, which enables OAM as a new form of multiple access technology \cite{Intro1}. Typical approaches to generate OAM waves including using helical phase
plate antenna \cite{Intro4}, helicoidal parabolic antenna \cite{Intro6}, and uniform circular array (UCA) antenna \cite{Intro7,Intro8,Intro9}. Due to the advantage of  digitally generating multiple OAM-modes simultaneously,
the UCA-based OAM is extensively studied \cite{Tunable_G,Beam_S}. Chen {\sl et al.} has designed a complete OAM-based point-to-point wireless communication scheme in the LoS scenario \cite{Intro3}, which shows that the UCA-based OAM transmission can achieve higher spectrum efficiency than multiple-input multiple-output orthogonal frequency division multiplexing (MIMO-OFDM).

To achieve the interference-free transmission in UCA-based OAM transmissions, the perfect alignments between transmitters and receivers are strictly required \cite{Joint_OAM_multiplexing}. But in real applications, it is hard to achieve ideal alignments, the phase turbulence, caused by misaligned transmitters and receivers, will lead to serious inter-mode interference, thereby the signals cannot be correctly detected at the receiver \cite{Performance_M}. 
To overcome this problem, Suganuma {\sl et al.} in \cite{Inter_mode_i} proposes an inter-mode interference suppression method that employs only even-numbered modes for UCA-based OAM multiplexing, the odd-numbered modes are unused to reduce the inter-mode interference from adjacent modes in the presence of the beam axis misalignment, which has lower complexity but the system capacity is also reduced. To avoid the capacity reduction,
\cite{Reception_of} further proposes a UCA-based multi-mode OAM reception method including the beam steering with the estimated angle of arrival (AoA)
and the amplitude detection with the estimated distance, simulation results shows that the proposed OAM method in \cite{Reception_of} can completely eliminate the effect of the misalignment error and approaches the performance of ideally aligned OAM channels. To further reduce the design complexity, Jing {\sl et al.} in \cite{Fast_T} proposes a fast transceiver design to reduce the impact of misaligned receivers in OAM communication systems.

Existing researches on UCA-based OAM communications mainly focus on the point-to-point coaxial transmissions in the LoS scenarios \cite{Inter_mode_i,Reception_of,Fast_T,Intro2,A_2_D}, i.e., the single-user OAM communications with LoS links. Few works have extended to multi-user OAM communications, since in multi-user OAM cases, it is impossible for the transmitter to be aligned with all the receivers, so the inter-mode interference becomes much more complicated to handle. In addition, the inter-user OAM interference among multiple receivers also needs to be carefully mitigated in order to improve the system performance. To address the above issues, Chen {\sl et al.} in \cite{Multi-user} utilizes a beam steering matrix to implement the diagonalization decomposition to eliminate the inter-mode interference under the assumption that the transmission distance is much larger than the radius of the UCA, thereby some specific phase parameters can be ignored to simplify the design. Besides, \cite{Precoding_b} designs a fractional programming based precoder design to maximize the sum capacity by using quadratic transformation, which has no requires for the system settings, but has higher design complexity. And \cite{Joint_spatial} proposes a multi-user OAM preprocessing scheme to eliminate the co-mode and inter-mode interferences in the downlink transmissions, so the signals can be directly decoded at the receiver side.

All the above multi-user OAM researches assume that the ideal LoS transmission paths are exist between the transmitter and receivers. However, this situation is difficult to guarantee in practical applications, due to the more and more complex wireless transmission environments \cite{Joint_beamforming}. Thereby, reconfigurable intelligent surfaces (RISs) can be introduced to proactively construct LoS links to guarantee the system performance \cite{Cooperative_beamforming, Joint_user_association}. Motivated by this, we investigate the joint precoder and reflector design for RIS-assisted multi-receiver OAM communication systems. We propose a three-layer designs at the transmitter side, two of them are used to cancel the inter-user OAM mode interference and the inter-mode self-interference in the multi-user OAM communication scenarios, 
 thereby we can transform the overall channel matrix into the equivalent circulant matrix for canceling the phase turbulence, and the last layer is used to achieve the power allocation among different users. Furthermore, we leverage RIS to overcome unfavorable propagation environments, i.e., creating LoS transmission links to improve the system performance. Finally, the precoder design and the RIS's reflection are jointly optimized to maximize the transmission capacity. The novelty and main contributions of this paper are summarized as follows:

	\begin{itemize}
		\item We build a RIS-assisted multi-user OAM communication system, where the channel matrices between the transmitter, the RIS and all the users are derived. Based on this, we formulate the sum rate maximization problem for multi-user OAM systems through the precoder design and the RIS's phase shift optimization.
	
	    \item For the precoder design, we propose a three-layer scheme, which includes inter-user OAM mode interference cancellation, inter-mode self-interference elimination and the power allocation. Besides, for the sub-matrices used to do the inter-user OAM mode interference cancellation and the inter-mode self-interference elimination, we directly give the specific  expressions by analyzing the characteristics of the overall channel, which can significantly reduce the complexity for system design. And finally the power allocation is achieved trough traditional water-filling methods. 
	    
	    \item With the proposed three-layer precoder design, we are able to transform the multi-user OAM transmission system into a virtual MIMO system with distributed receiving antennas. By further analyzing the overall channel model, we are able to obtain the closed-form expressions for updating the RIS's phase shift design, which can avoid the high optimization complexity caused by the large number of reflecting elements at the RIS.
	
		\item We compare the proposed multi-user OAM communication scheme with traditional MIMO transmission schemes, numerical results have shown that our proposed multi-user OAM transmission scheme can achieve better sum rate performance due to the well-designed orthogonality among different receivers and OAM modes.
		
	\end{itemize}

	The rest of the paper is organized as follows. In section II, we introduce the RIS-assisted multi-user OAM communication systems, the channel models and problem formulations are also included in section II. The proposed three-layer precoder design for inter-user OAM mode interference cancellation and inter-mode self-interference elimination are introduced in section III. In section IV, we explain the detailed processes for RIS's phase shift design. The performance evaluations are provided in section V. And finally, we conclude this work in section VI.
	
	The notations are listed as follows.
	Bold symbols in capital letter and small letter denote matrices and vectors, respectively.
	$\mathcal{CN}(\mu,\sigma^{2})$ denotes the circularly symmetric complex Gaussian (CSCG) distribution with mean $\mu$ and variance $\sigma^{2}$.
	$\mathbf{A}=\diag(\bm{a})$ means that $\mathbf{A}$ is the diagonal matrix of the vector $\bm{a}$. $A_{i,j}$ represents the element at row $i$, column $j$ of matrix $\mathbf{A} $ and $a_{n}$ represents the $n$-th elements of vector $\bm{a}$. $\det(\mathbf{A})$ means the determinant of matrix $\mathbf{A}$.
	$\otimes$ is the Kronecker product, while $\bm{a}\times \bm{b}$ represents the cross product of vector $\bm{a}$ and $\bm{b}$.
	$||\mathbf{w}||$ denotes the Euclidean norm.
	$\mathbf{G}^{T}$ and $\mathbf{G}^{H}$ denote the transpose and conjugate transpose of matrix $\mathbf{G}$, respectively. $\bm{1}_M$ is the $M \times M$ identity matrix.

	
\begin{figure}[t]
	\centering
	\includegraphics[width=0.45\textwidth]{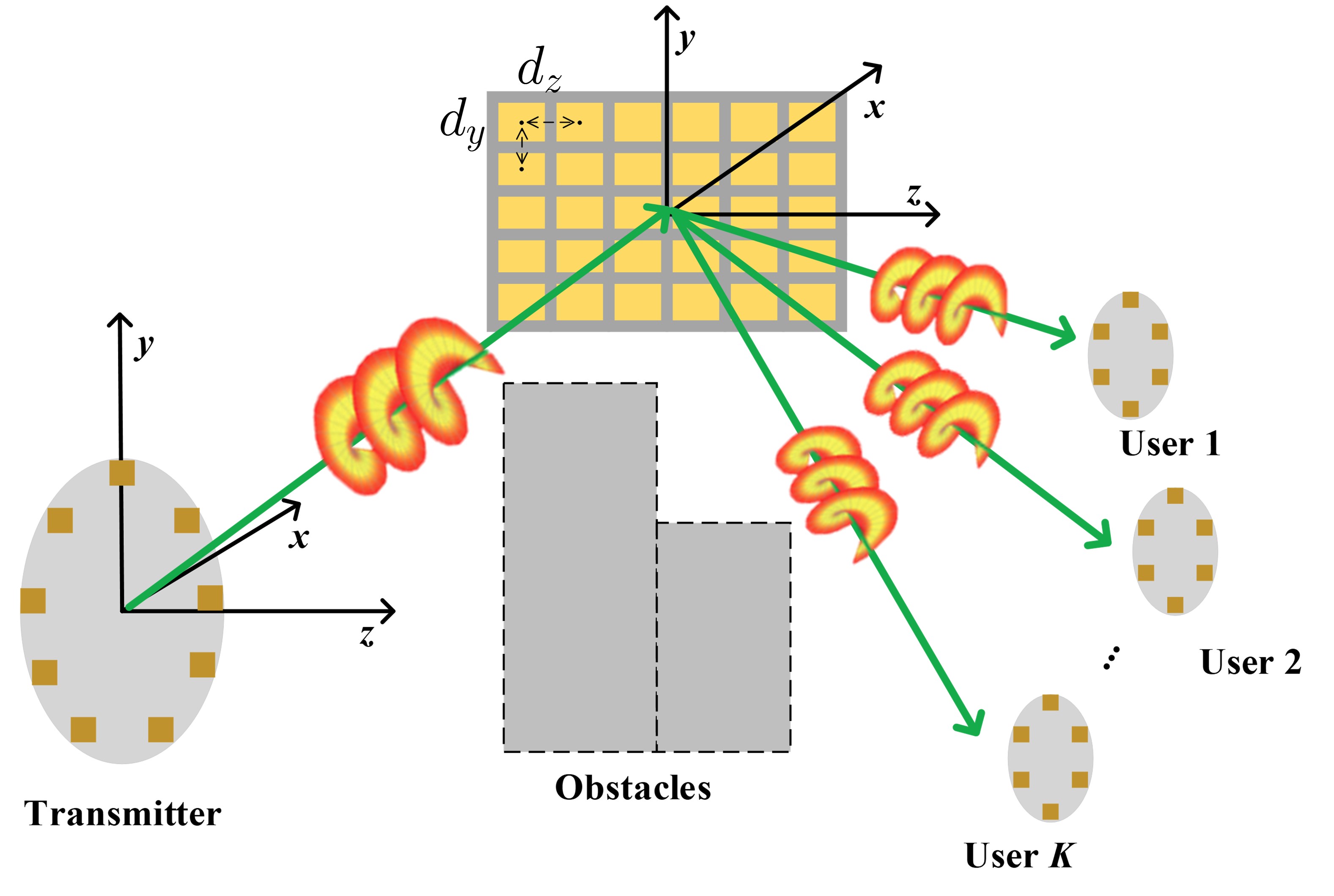}\\
	\caption{The considered RIS-assisted multi-user OAM communication system.}
	\label{system_model}
\end{figure}	
	\section{RIS-assisted Multi-user OAM Wireless Communication Systems }
	Employing UCA is a propular method to generate and receive OAM beams due to the simple structure and multi-mode multiplexing ability \cite{Tunable_G,Beam_S}. In this paper, we consider a UCA-based downlink multi-user OAM communication system, where multi-mode OAM beams are generated by a UCA-based transmitter with $N_T$ transmitting antennas. And $K$ users are randomly located in the area, each composed with $N_R$-element UCA as shown in Fig. \ref{system_model}. And we assume that the direct transmission links between the transmitter and the users are blocked by obstacles, thereby a RIS with $M$ reflecting elements is adopted to crate LoS transmission links in the system. 	
	Moreover, for the $N_T$-element transmit UCAs, at most $N_T$ OAM modes can be generated\cite{Intro7}. In the considered downlink OAM multi-user communication systems, the transmitter is required to generate $K$ OAM modes to carry signals for $K$ users, each equipped with $N_R$ receiving antennas at the same time. Owing to the limitation on the number of multiplexed OAM modes,
	the requirements $N_T \geq KN_R$ should be satisfied
	 to guarantee that all modes can be successfully separated at the user sides. Without loss of generality, we assume $N_T=KN_R$ in this paper  \cite{Multi-user,Joint_spatial, Achievable_rate_maximization}.  
	
\subsection{Channel Model for RIS-assisted Multi-user OAM 
	 Communication Systems}
		\begin{figure}[t]
	\centering
	\includegraphics[width=0.45\textwidth]{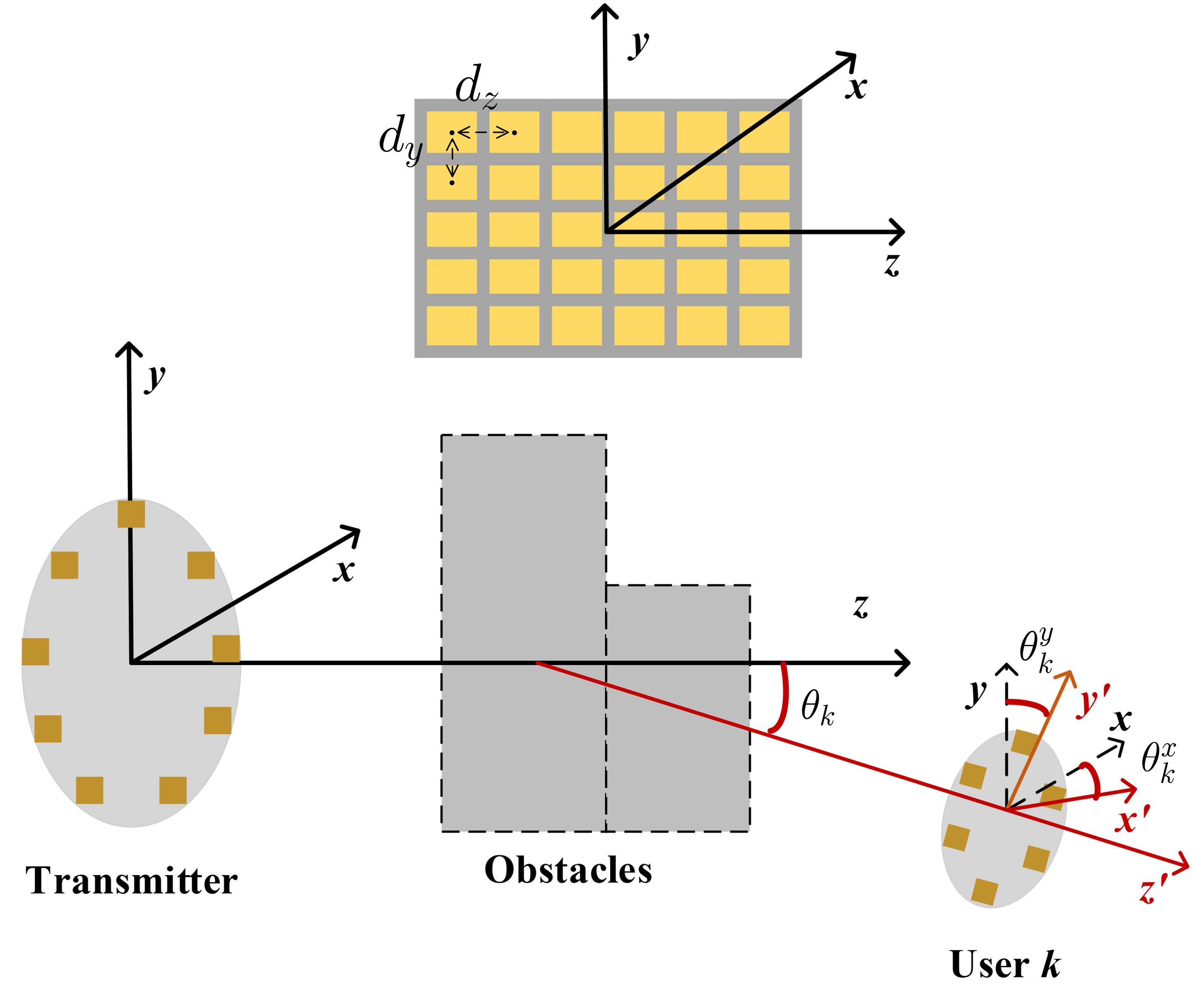}\\
	\caption{Channel model for the considered multi-user OAM communication system.}
	\label{channel_model}
\end{figure}	
The incident channel from the transmit UCA to the RIS, the reflected channel from the RIS to the user $k$, $k=1,2,...,K$, are denoted by $\mathbf{H} \in  \mathbb{C}^{M \times N_T}$ and $\mathbf{G}_{k} \in \mathbb{C}^{N_R \times M}$, respectively. 
All the channels can be calculated according to the geometrical distribution of the transmitter, RIS and users. To make it easier to illustrate, we use one user to explain detaily as shown in Fig. \ref{channel_model}. The center of the transmit UCA is located at the origin of the coordinate while the center of the user $k$ is located at $(x_k, y_k, z_k)$. $\theta_k^x$ (the angle between axes $x$ and $x'$) and $\theta_k^y$ (the angle between axes $y$ and $y'$) denote the angle between the transmit and the user UCAs’ normal lines along the $x$ and $y$ axes, respectively, with the constraints $0 \leq \theta_k^x \leq \pi/2$ , $0 \leq \theta_k^y \leq \pi/2$. The deflection angle between the transmit and the user $k$' normal lines, denoted by $\theta_k$, can be given as $\theta_k =\arctan \sqrt{(\tan^2 \theta_k^x+\tan^2 \theta_k^y)}$. With the above defination, the coordinates of the $n$-th transmit element and $l$-th receive element at user $k$ can be expressed as $\bm{a}_{T,n}$ and $\bm{a}_{k,l}$, which are given in the following \cite{Achievable_rate_maximization}
	\begin{equation}	 
	\begin{aligned}  \label{location_T}  
	 \bm{a}_{T,n}=\left( R_t \cos \left(\frac{2 \pi (n-1)}{N}\right),       R_t \sin \left(\frac{2 \pi (n-1)}{N}\right),0                         \right), 
	\end{aligned}
	\end{equation}	 

	\begin{equation}	  
		\begin{aligned} \label{location_k} 
	\bm{a}_{k,l} = (x_k, y_k, z_k) &- R_r\cos \left( \frac{2\pi(l-1)}{L}\right)
	\frac{\bm{b}^{k}}{||\bm{b}^{k}||}\\ &+ 
	R_r \sin \left( \frac{2\pi(l-1)}{L}\right)\frac{\bm{c}^{k}}{||\bm{c}^{k}||}.
	\end{aligned}
\end{equation}
where $R_t$ is the radius of transmit UCA, and $R_r$ is the radius of user $k$' receive UCA, here we assume that all the user has the same size of receiving antennas. 
The expressions of parameters $\bm{b}^{k}$ and $\bm{c}^{k}$ are shown in the following:
	\begin{equation}
	\begin{aligned} \label{L3}
		\bm{b}^{k} = [\tan \theta_k^x, \tan \theta_k^y,1] \times [1,0,0],
	\end{aligned}
\end{equation}
\begin{equation}
	\begin{aligned} \label{L4}
		\bm{c}^{k} = [\tan \theta_k^x, \tan \theta_k^y,1] \times \bm{b}^{k}.
	\end{aligned}
\end{equation}
Here, $\times$ represents the cross product operation of two three-dimensional vectors.
In this paper, we use the uniform planar array (UPA) as the RIS, each row of the RIS has $M_z$ elements and each column has $M_y$ elements, then we have $M=M_z M_y$. The distance between the centers of any two reflecting elements along the $y$ and $z$ axis are $d_y$ and $d_z$, respectively.
we assume the center of RIS is located at $(x_R, y_R, z_R)$. Then the coordinates of all the passive reflecting elements on the RIS can be expressed as
\begin{equation}	 
	\begin{aligned}  \label{location_RIS}
		&\mathbf{A}_{RIS}= [x_R, y_R, z_R]^T\otimes \mathbf{1}_M^T
		+ \\  &\left[
		\mathbf{0}_M, 
		d_y\left(\!\!\bm{m}_y \!\otimes \!\mathbf{1}_{M_z} \!+\! \frac{1\!-\!M_y}{2}\!\!\right),
		d_z\left(\!\!\mathbf{1}_{M_y} \!\otimes \!\bm{m}_z \!+\! \frac{1\!-\!M_z}{2}\!\!\right)
		\right]^T.
	\end{aligned}
\end{equation}	

In equation \eqref{location_RIS}, $\mathbf{A}_{RIS}$ is a $3 \times M $ dimensional matrix, with each column $\bm{a}_{R,m}$ represents the 3D location of the $m$-th element.
Besides, $\bm{m}_y=[0,1,...,M_y-1]^T$ and $\bm{m}_z=[0,1,...,M_z-1]^T$. 
Given the above definitions, we can obtain $(m,n)$-th element of channel $\mathbf{H}$ from transmit UCA to the RIS, the $(l,m)$-th element of reflected channel $\mathbf{G}_{k}$ from the RIS to the user $k$ as \cite{Joint_reflection,Quasi-fractal}
\begin{equation}
	\begin{aligned} \label{element1}
	H(m,n)= \beta \frac{\lambda}{4\pi||\bm{a}_{R,m}-\bm{a}_{T,n}||}e^{-j \frac{2\pi}{\lambda}||\bm{a}_{R,m}-\bm{a}_{T,n}||}	,
	\end{aligned}
\end{equation}
\begin{equation}
	\begin{aligned} \label{element2}
		G_k(l,m)= \beta \frac{\lambda}{4\pi||\bm{a}_{k,l}-\bm{a}_{R,m}||}e^{\frac{-j2\pi}{\lambda}||\bm{a}_{k,l}-\bm{a}_{R,m}||},
	\end{aligned}
\end{equation}
where $\beta$ denotes the parameter gathering relevant constants on antenna array-elements, and $\lambda$ represents the carrier wavelength.

The phase shift matrix at the RIS can be expressed as $\mathbf{\Phi} \in \mathbb{C}^{M \times M}$, which can be further expressed as $\mathbf{\Phi}= \diag (\pmb{\phi})$. $\pmb{\phi}=[\phi_{1},\phi_{2},...,\phi_{M}]^T$ is the corresponding reflection vector of the RIS, where
$\phi_{m}=\beta_{m}e^{j\theta_{m}}$, $m = 1,2,...,M$. Moreover, $\theta_{m} \in [0,2\pi)$ and $\beta_{m} \in [0,1]$ respectively represent the phase and amplitude change brought by the $m$-th reflecting element of RIS to the incident signals.
Without loss of generality, we set $\beta_{m} =1$, which means that all the elements of the RIS are switched on to fully reflect the incident signals.

\subsection{Signal Model and Problem Formulation for RIS-assisted Multi-user OAM Communication Systems }

\begin{figure}[t]
	\centering
	\includegraphics[width=0.45\textwidth]{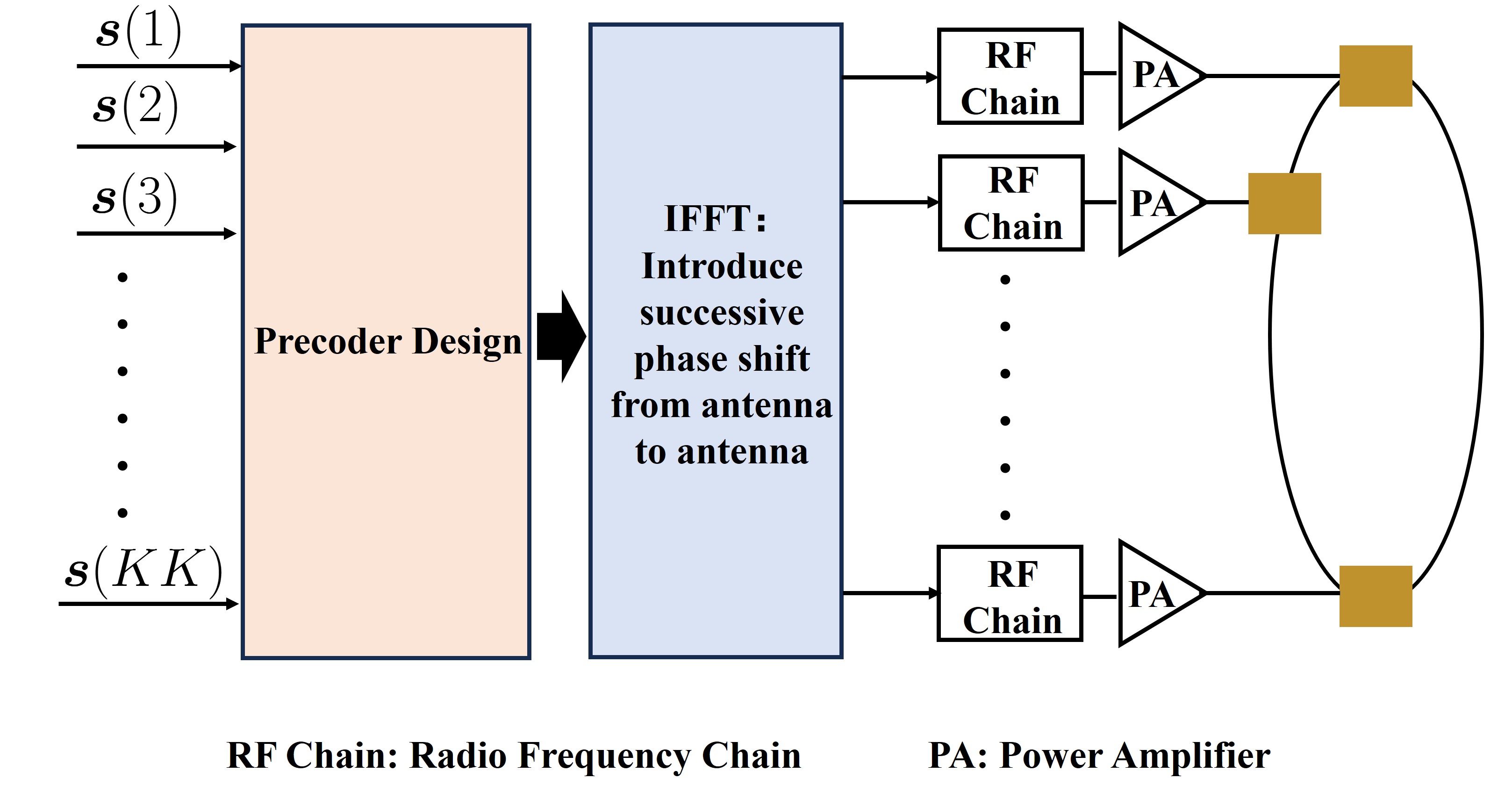}\\
	\caption{The transmitter structure of the proposed multi-user OAM communication system.}
	\label{Transmit_Structure}
\end{figure}	
 \vspace{0.5cm}
\begin{figure}[t]
	\centering
	\includegraphics[width=0.45\textwidth]{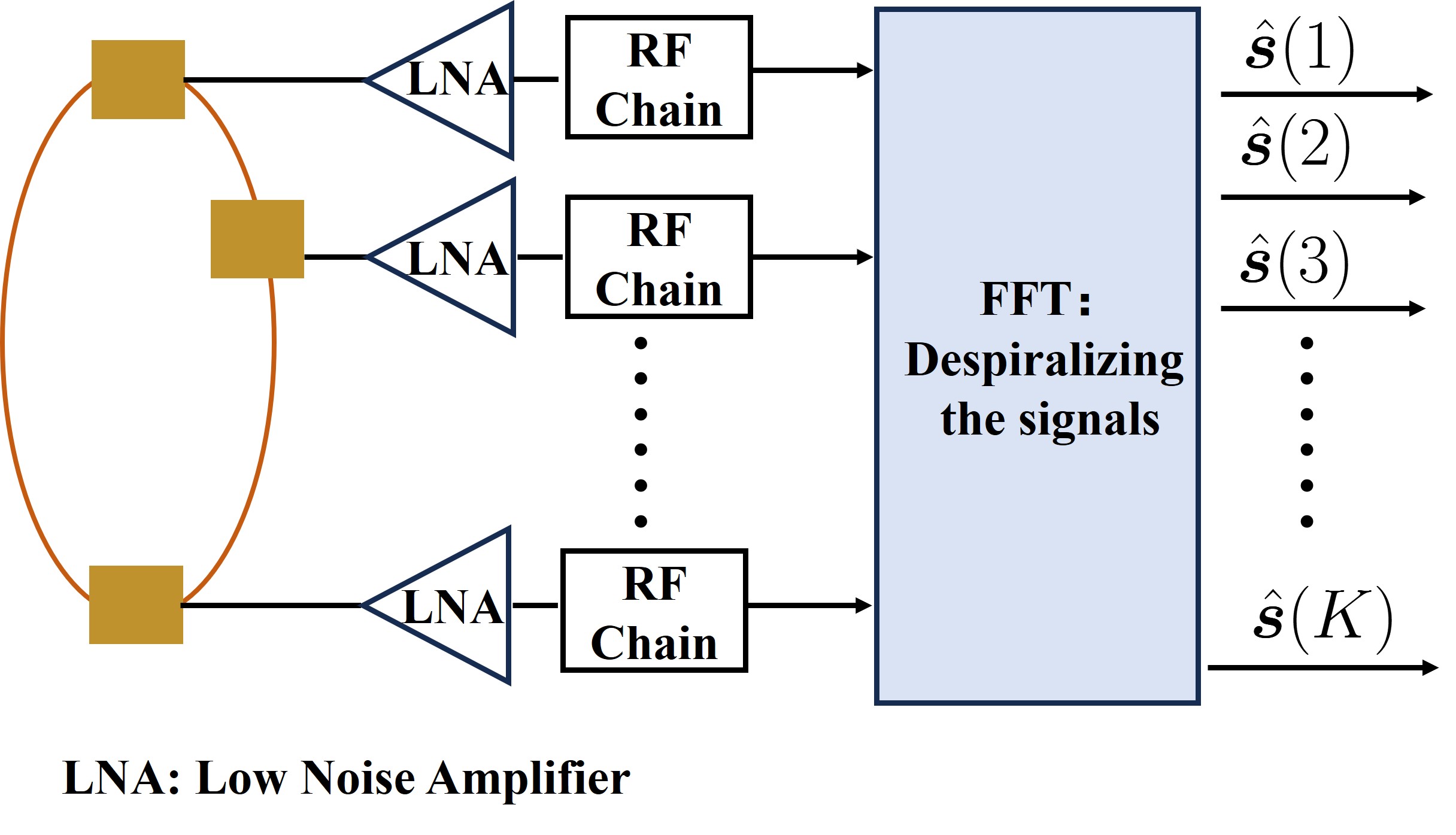}\\
	\caption{The receiver structure at the user side of the proposed multi-user OAM communication system.}
	\label{Receiver_Structure}
\end{figure}	
The OAM beams are generated by the transmit UCA by feeding its antenna elements 
with a successive phase shift \cite{PhysRevA.45.8185}, and the baseband analog inverse fast fourier transform (IFFT) is used at the transmitter side to induce the successive phase shifts as shown in Fig. \ref{Transmit_Structure}. The equivalent baseband signal model of multi-user OAM data symbols can be expressed as $\mathbf{F}\bm{s} $, where $\mathbf{F}=\mathbf{I}_K \otimes \mathbf{F}_K$ represents the $K$-dimensional
 block diagonal matrix used to simultaneously generate multi-mode
 OAM beams, where $\mathbf{I}_K$ donates the $K \times K$ dimensional identity matrix, $\otimes$ is the Kronecker product, and
 $\mathbf{F}_K=\left[  \bm{f}(1)^H,  \bm{f}(2)^H,..., \bm{f}(K)^H \right] \in \mathbb{C}^{N_R \times K}$ is a
 $N_R \times K$ (partial) IFFT matrix with $\bm{f}(k)=[1, e^{-j\frac{2\pi k}{N_R}},...,e^{-j\frac{2\pi k(N_R-1)}{N_R}}]$ \cite{Multi-user,Joint_spatial}. The signal vector $\bm{s}$ contains all the data symbols that will be transmitted to $K$ users, and can be further writted as $\bm{s}^T=[\bm{s}_1^T,\bm{s}_2^T,...,\bm{s}_K^T] \in \mathbb{C}^{1 \times KK}$ with $\bm{s}_k^T=[s_k(1),s_k(2),...,s_k(K)] \in \mathbb{C}^{1 \times K}$. Besides, since the UCA at the transmitter and the UCAs at $K$ users are misaligned, the signals need to be precoded before IFFT processing, i.e., $\mathbf{F}\mathbf{W}\bm{s} $, where $\mathbf{W} \in \mathbb{C}^{KK \times KK}$ is the matrix for precoder design that used to eliminate the interferce at the transmitter, which can be further expressed as 
  $\mathbf{W}=\left[\mathbf{W}_1,\mathbf{W}_2, ..., \mathbf{W}_K   \right]$, with $\mathbf{W}_k$, $k=1,2,...,K$, representing the precoding matrix for user $k$.

 After going through the wireless channel, the received signals at user $k$ can be written as 
 
 \begin{equation}
 	\begin{aligned} \label{received_signal}
 		\bm{y}(k)=\mathbf{G}_k\mathbf{\Phi}\mathbf{H}\mathbf{F}\mathbf{W}\bm{s}+\bm{n}_k,
 	\end{aligned}
 \end{equation}
 where $\bm{n}_k$ represents the noise at the user $k$.

  The UCA-based OAM
 receiver at the user side has the similar baseband digital structure to the transmitter \cite{A_2_D}, but with the opposite
 operations, i.e., separating different OAM modes and despiralizing
 each mode by doing the analogy FFT operation as shown in Fig. \ref{Receiver_Structure}. Thus, the detected OAM data symbols at the user $k$ can be expressed as 
 
  \begin{equation}\label{k_re}
 	\begin{aligned} 
\bm{x}(k)=\mathbf{F}_{K}^H 	\mathbf{G}_k\mathbf{\Phi}\mathbf{H}\mathbf{F}\mathbf{W}\bm{s} + \mathbf{F}_{k}^H	\bm{n}_k.
 	\end{aligned}
 \end{equation}
 
Based on equation \eqref{k_re}, the signal-to-interference-plus-noise ratio (SINR) at user $k$ can be expressed as 

 \begin{equation}
 	\begin{aligned} \label{SINR_k}
 		\gamma_k = \frac{||\mathbf{F}_{K}^H 	\mathbf{G}_k\mathbf{\Phi}\mathbf{H}\mathbf{F}\mathbf{W}_k||^2}
 		{\sum_{i=1,i\neq k}^K||\mathbf{F}_{K}^H 	\mathbf{G}_k\mathbf{\Phi}\mathbf{H}\mathbf{F}\mathbf{W}_i||^2 + \sigma^2}.
 	\end{aligned}
 \end{equation}
 
 Our design goal is to maximize the capacity through the
 inter-user OAM mode interference cancellation, the
  inter-mode self-interference elimination and RIS's reflection design, under the transmit power constraint and RIS's phase shift constraint. The optimization problem can be shown as follow 
  
 \begin{align}
 	\textbf{(P1)} \quad \quad &\max_{\mathbf{W},\mathbf{\Phi}} \quad C=\sum_{k=1}^K \log(1+\gamma_k),\label{OBJ}	\\
 	 & \quad  s.t. \quad \sum_{k=1}^{K}||\mathbf{W}_{k}||^{2}\leq P_{T}, \label{PC} \\
 	 & \quad \quad \quad \quad |\phi_{m}|=1, \quad m=1,2,...,M. \label{PSC}
 \end{align}
 
Compared with traditional MIMO transmission systems, the channel model for multi-user OAM communication systems are much more complicated, which increases the difficulties of system analysis. Besides, during the precoder design, we need to cancel the inter-mode self-interference caused by the misalignment between the transmitter and the receivers, while mitigate the inter-user OAM mode interference due to the simultaneous transmissions by all the users, which need us to fully exploit the channel characteristics of multi-user OAM channels. Furthermore, since the users in considered systems also have multiple antennas, while most of the existing designs are based on single-antennas receiver, which makes most existing solutions unable to solve problem \textbf{(P1)}. In the following, by carefully analyzing the channel characteristics for multi-user OAM systems, we are able to provide a low complexity precoder design for inter-mode self-interference cancellation and inter-user OAM mode interference elimination, and the closed-from expression for updating RIS's phase shift is also given by further transforming the objective function. Detailed information for the precoder and RIS's designs are given in the following section.

\section{Precoder Design for Multi-user OAM Communication Systems } \label{Section11}
In considered multi-user OAM communication systems, the transmitter and receivers are misaligned, which leading to severe inter-mode self-interference. Besides, all the signals for $K$ users are transmitted 
simultaneously, thereby we need effective inter-user interference mitigation design to improve the system performance. Thereby, the main propose of the precoder design is to eliminate the inter-mode self-interference and the inter-user OAM mode interference, so the signals can be successfully separated at the user side.
First, we need to characterize the equivalent channels.
With equation \eqref{k_re}, the equivalent channel $\mathbf{\Gamma}(k) \in \mathbb{C}^{K \times KK}$ from transmitter to user $k$ can be modeled as 

\begin{equation}
	\begin{aligned} \label{overall_channel}
		\mathbf{\Gamma}(k)= \mathbf{F}_K^H\mathbf{G}_k\mathbf{\Phi} \mathbf{H}\mathbf{F},
	\end{aligned}
\end{equation}

which includes the IFFT and FFT operations at the transmitter and user sides, respectively.
	Then the overall channel $\mathbf{\Gamma} \in \mathbb{C}^{KK \times KK}$ from transmitter to all the users can be expressed as 
	
	\begin{equation}
		\begin{aligned} \label{alluser_channel}
\mathbf{\Gamma}= 
	\left[
\begin{matrix}
\mathbf{F}_K^H\mathbf{G}_1\\
\mathbf{F}_K^H\mathbf{G}_2\\
\vdots\\
\mathbf{F}_K^H\mathbf{G}_K
\end{matrix}
\right] \mathbf{\Phi}\mathbf{H}\mathbf{F} = 
\mathbf{B} \mathbf{\Phi}\mathbf{H}\mathbf{F}=
\left[
\begin{matrix}
	\mathbf{\Gamma}_1\\
	\mathbf{\Gamma}_2\\
	\vdots\\
	\mathbf{\Gamma}_K
\end{matrix}
\right],
		\end{aligned}
	\end{equation}
	
where $\mathbf{B}=\left[\mathbf{F}_K^H\mathbf{G}_1,
\mathbf{F}_K^H\mathbf{G}_2,\cdots,\mathbf{F}_K^H\mathbf{G}_K
\right]^T$ is a $KK\times M$ dimensional matrix, and $\mathbf{\Gamma}_k$, $k=1,2,...,K$, represents the $k$-th row of the overall channel $\mathbf{\Gamma}$.
Then the overall channel $\mathbf{\Gamma}$ is multiplied with the precoding matrix $\mathbf{W}$ to mitigate the inter-mode self-interference and the inter-user OAM mode interference, which can be expressed as 
\begin{equation}
	\begin{aligned} \label{111}
		\mathbf{\Gamma}\mathbf{W}&=\left[
		\begin{matrix}
			\mathbf{\Gamma}_1\\
			\mathbf{\Gamma}_2\\
			\vdots\\
			\mathbf{\Gamma}_K
		\end{matrix}
		\right]
		\left[
		\begin{matrix}
			\mathbf{W}_1, \mathbf{W}_2,
			\dots,
			\mathbf{W}_K
		\end{matrix}
		\right]\\
		&=
		\left[
		\begin{matrix}
\mathbf{\Gamma}_1\mathbf{W}_1 &\mathbf{\Gamma}_1\mathbf{W}_2
			&\dots
			&\mathbf{\Gamma}_1\mathbf{W}_K     \\
\mathbf{\Gamma}_2\mathbf{W}_1 &\mathbf{\Gamma}_2\mathbf{W}_2
			&\dots
			&\mathbf{\Gamma}_2\mathbf{W}_K      \\
\vdots  &\vdots &   \ddots &\vdots \\
\mathbf{\Gamma}_K\mathbf{W}_1 &\mathbf{\Gamma}_K\mathbf{W}_2
&\dots
&\mathbf{\Gamma}_K\mathbf{W}_K 
		\end{matrix}
		\right].
	\end{aligned}
\end{equation}

For the precoder design $\mathbf{W}$, there are three requirements:	

\begin{itemize}
	\item  \textbf{Requirement 1}: The overall channel matrix and the precoding matrix need to satisfy $\mathbf{\Gamma}_i\mathbf{W}_j=0$, $\forall i\neq j$. This constraint guarantees that the inter-user OAM mode interference, i.e., the interference among different modes at different users, is eliminated.
	
	\item \textbf{Requirement 2}:
	For the sub-matrix $\mathbf{\Gamma}_i\mathbf{W}_i$ on the diagonal in equation \eqref{111}, all the non-diagonal elements in  $\mathbf{\Gamma}_i\mathbf{W}_i$ are require to be removed, so that $\mathbf{\Gamma}_i\mathbf{W}_i$ can be diagonalized, which will cancel the inter-mode self-interference of each user.
	
	\item \textbf{Requirement 3}:
	After handling the inter-user and inter-mode interference, the overall transmit power at the transmitter should be carefully allocate to all the user to improve the sum rate capacity of the system.
	\end{itemize}

To meet the above requirements, we propose a three-layer precoder design, which divides $\mathbf{W}$ into three sub-precoding matrix as $\mathbf{W}=\mathbf{Q}\mathbf{D}\mathbf{E}$, where sub-precoding matrix $\mathbf{Q}$ is used for inter-user OAM mode interference cancellation, sub-precoding matrix $\mathbf{D}$ is responsible for inter-mode self-interference elimination, and sub-precoding matrix $\mathbf{E}$ is for transmit power allocation at the transmitter side.
In the following, we will give the specific processes about how to derive each sub-matrix and how to get 
the overall precoding matrix.

\subsection{Pre-processing for Inter-user Interference Cancellation}
First we start with the inter-user interference, since the inter-user interference mitigation is a common and important problem in all kinds of multi-user systems. To handle the inter-user interference,  
We need to construct a sub-precoding matrix as shown in the following 
\begin{equation}
	\begin{aligned} \label{inter_user}
		\mathbf{Q}= \left[\mathbf{Q}_1, \mathbf{Q}_2,..., \mathbf{Q}_K\right],
	\end{aligned}
\end{equation}
which is a $KK \times KK$ dimensional matrix with $\mathbf{Q}_k \in \mathbb{C} ^{KK \times K}$. To meet the first requirement, 
$\mathbf{\Gamma}_k\mathbf{Q}_i=0$, $\forall k\neq i$ should be satisfied, which means that $\mathbf{Q}_i$ should be select from the null space of $\bar{\mathbf{\Gamma}}_k$ \cite{Joint_spatial}, and $\bar{\mathbf{\Gamma}}_k$ is
composed of the channel consisting of all other receivers except user $k$, and can be expressed as
\begin{equation}
	\begin{aligned} \label{null_space}
	\bar{\mathbf{\Gamma}}_k= 
		\left[
		\begin{matrix}
			\mathbf{\Gamma}_1,
			\mathbf{\Gamma}_2,
			\cdots,
			\mathbf{\Gamma}_{k-1},
			\mathbf{\Gamma}_{k+1},
			\cdots,
			\mathbf{\Gamma}_K
		\end{matrix}
		\right]^T.
	\end{aligned}
\end{equation}
The null-space of $\bar{\mathbf{\Gamma}}_k$ can be calculated by computing the Singular Value Decomposition (SVD) of $\bar{\mathbf{\Gamma}}_k$, which can be further expressed as
\begin{equation}
	\begin{aligned} \label{decom}
	\bar{\mathbf{\Gamma}}_k &= \bar{\mathbf{U}}_k \bar{\mathbf{\Sigma}}_k \bar{\mathbf{V}}_k
	=\bar{\mathbf{U}}_k \bar{\mathbf{\Sigma}}_k 
	\left[ \bar{\mathbf{V}}_k^{(1)},    \bar{\mathbf{V}}_k^{(2)} \right].
	\end{aligned}
\end{equation}
In equation \eqref{decom}, $\bar{\mathbf{U}}_k$ is a $K(K-1) \times K(K-1)$ dimensional left singular matrix, $\bar{\mathbf{\Sigma}}_k$ is a $K(K-1) \times KK$ matrix including all singular values of $\bar{\mathbf{\Gamma}}_k$, and $\bar{\mathbf{V}}_k$ is the $KK \times KK$ right singular matrix. Denote the rank of $\bar{\mathbf{\Gamma}}_k $ as $\eta_k$, then $\bar{\mathbf{V}}_k$ can be divided into two parts $\bar{\mathbf{V}}_k^{(1)}$ and $\bar{\mathbf{V}}_k^{(2)}$, where $\bar{\mathbf{V}}_k^{(1)}$ holds the first $\eta_k$ singular vectors, and $\bar{\mathbf{V}}_k^{(2)}$ holds the last $(KK-\eta_k)$ right singular vectors.  Therefore, $\bar{\mathbf{V}}_k^{(2)}$ forms an orthogonal basis for the null space of $\bar{\mathbf{\Gamma}}_k$. As $\eta_k \leq (K-1)K$, $\bar{\mathbf{V}}_k^{(2)}$ contains at least $K$ sight singular vectors. Therefore, we choose the last $K$ columns of $\bar{\mathbf{V}}_k^{(2)}$ to form $\mathbf{Q}_k$. 

After eliminating the inter-user interference by the first layer precoding matrix $\mathbf{Q}$, the equivalent channel from transmitter to all the users becomes 

\begin{equation}
	\begin{aligned} \label{efective_chann}
		\mathbf{\Gamma}\mathbf{W}&=\left[
		\begin{matrix}
			\mathbf{\Gamma}_1\\
			\mathbf{\Gamma}_2\\
			\vdots\\
			\mathbf{\Gamma}_K
		\end{matrix}
		\right]
		\left[
		\begin{matrix}
			\mathbf{Q}_1, \mathbf{Q}_2,
			\dots,
			\mathbf{Q}_K
		\end{matrix}
		\right]\\
&		=
	\left[
	\begin{matrix}
		\mathbf{\Gamma}_1\mathbf{Q}_1 &\mathbf{0}
		&\dots
		&\mathbf{0}     \\
		\mathbf{0} &\mathbf{\Gamma}_2\mathbf{Q}_2
		&\dots
		&\mathbf{0}      \\
		\vdots  &\vdots &   \ddots &\vdots \\
		\mathbf{0} &\mathbf{0}
		&\dots
		&\mathbf{\Gamma}_K\mathbf{Q}_K 
	\end{matrix}
	\right].
	\end{aligned}
\end{equation}
where $\mathbf{\Gamma}_k\mathbf{Q}_k$, $k=1,2,...,K$ represents the effective channel from transmitter to receiver $k$. Compared with equation \eqref{11}, we can see that the overall channel is decomposed into $K$ parallel single-user OAM channels by multiplying 
$\mathbf{Q}$.

\vspace{-0.2cm}

\subsection{Pre-processing for Inter-mode Interference Elimination}
For OAM communication systems, the transmitter and receiver need to be perfectly aligned to guarantee the successfully decoding of the received signals. However, in multi-user OAM systems, perfect alignment between the transmitter and all the receivers is impossible.
The angles $\theta_k$ between the transmitter and each receiver $k$ causes inter-mode self-interference, which cause $\mathbf{\Gamma}_k\mathbf{Q}_k$ not being a diagonal matrix \cite{Intro3}. Thus the purpose of inter-mode self-interference elimination is to remove the non-diagonal parameters in $\mathbf{\Gamma}_k\mathbf{Q}_k$. To meet this requirements, we construct the inter-mode self-interference elimination matrix as
\begin{equation}
	\begin{aligned} \label{inter_user}
		\mathbf{D}= \diag\left[
		\mathbf{D}_1, \mathbf{D}_2,...,\mathbf{D}_K
		\right],
	\end{aligned}
\end{equation}
with $\mathbf{D}_k$, $k=1,2,...,K$ being the $K \times K$ dimensional inter-mode self-interference elimination matrix for user $k$.
By multiplying $\mathbf{D}$, the equivalent channel can be further reformulated as 
\begin{equation}
	\begin{aligned} \label{efective_chann1}
		\mathbf{\Gamma}\mathbf{W}&=\left[
		\begin{matrix}
			\mathbf{\Gamma}_1\\
			\mathbf{\Gamma}_2\\
			\vdots\\
			\mathbf{\Gamma}_K
		\end{matrix}
		\right]
		\left[
		\begin{matrix}
			\mathbf{Q}_1, \mathbf{Q}_2,
			\dots,
			\mathbf{Q}_K
		\end{matrix}
		\right]
	\left[
		\begin{matrix}
		\mathbf{D}_1 & &  &\\
		&      \mathbf{D}_2 & & \\
		& &  \ddots & \\
		& & & \mathbf{D}_K
	    \end{matrix}
	\right]
		\\
		&=
		\left[
		\begin{matrix}
			\mathbf{\Gamma}_1\mathbf{Q}_1\mathbf{D}_1 &\mathbf{0}
			&\dots
			&\mathbf{0}     \\
			\mathbf{0} &\mathbf{\Gamma}_2\mathbf{Q}_2\mathbf{D}_2
			&\dots
			&\mathbf{0}      \\
			\vdots  &\vdots &   \ddots &\vdots \\
			\mathbf{0} &\mathbf{0}
			&\dots
			&\mathbf{\Gamma}_K\mathbf{Q}_K \mathbf{D}_K
		\end{matrix}
		\right].
	\end{aligned}
\end{equation}
 
 The goal of $\mathbf{D}_k$ is to diagonalize the product , therefore, $\mathbf{D}_k$ can be easily formed by the inverse of $\mathbf{\Gamma}_k\mathbf{Q}_k$, i.e.,  $\mathbf{D}_k= \left(\mathbf{\Gamma}_k\mathbf{Q}_k \right)^{-1}$. With this operation, we can guarantee that $\mathbf{\Gamma}_k\mathbf{Q}_k\mathbf{D}_k$ satisfies the \textbf{Requirement 2}, but the diagonal element of $\mathbf{\Gamma}_k\mathbf{Q}_k\mathbf{D}_k$ is also normalized. To avoid this problem, we further adopt the SVD of $\mathbf{\Gamma}_k\mathbf{Q}_k$, which can be further expressed as $\mathbf{\Gamma}_k\mathbf{Q}_k=\tilde{\mathbf{U}}_k\tilde{\mathbf{\Sigma}}_k\tilde{\mathbf{V}}_k$, with $\tilde{\mathbf{\Sigma}}_k$ being the $K \times K$ dimensional diagonal matrix, and each value in the diagonal represents the singular value of $\mathbf{\Gamma}_k\mathbf{Q}_k$. By further multiply $\left(\mathbf{\Gamma}_k\mathbf{Q}_k \right)^{-1}$ with $\tilde{\mathbf{\Sigma}}_k$, we have the final inter-mode self-interference elimination matrix as shown in the following
  \begin{equation}
 	\begin{aligned} \label{inter-mode}
 		\mathbf{D}_k= \left(\mathbf{\Gamma}_k\mathbf{Q}_k \right)^{-1}\tilde{\mathbf{\Sigma}}_k.
 	\end{aligned}
 \end{equation}
With equation \eqref{inter-mode}, the diagonal elements of $\mathbf{\Gamma}_k\mathbf{Q}_k\mathbf{D}_k$, $k=1,2,...,K$ are recovered to the corresponding singular values, which benefits the following power allocation at the transmitter.

\subsection{Preprocessing for Power Allocation at Transmitter}
With the preprocessing matrices $\mathbf{Q}$ and $\mathbf{D}$, the inter-user OAM mode interference and the inter-mode self-interference can be completely eliminated. Thereby the SINR in equation \eqref{SINR_k} is reduced to signal-to-noise ratio (SNR) as shown in the following:
 
 \begin{equation}
	\begin{aligned} \label{SNR_k}
		\delta_k = \frac{||\mathbf{\Gamma}_k\mathbf{Q}_k \mathbf{D}_k \mathbf{E}_k||^2}
		{\sigma^2},
	\end{aligned}
\end{equation}
where $\mathbf{E}_k$ is the power allocation matrix at the transmitter side with $\sum_{k=1}^{K}|| \mathbf{E}_k|| \leq P_T$. And the original optimization problem \textbf{(P1)} is reduced to problem \textbf{(P2)} as shown in the follow:
\begin{align}
	\textbf{(P2)} \quad \quad &\max_{\mathbf{E}_k,\mathbf{\Phi}} \quad C_k=\sum_{k=1}^K \log(1+\delta_k),	\\
	& \quad  s.t. \quad \sum_{k=1}^{K}||\bm{E}_{k}||^{2}\leq P_{T}, \label{PC1} \\
	& \quad \quad \quad \quad |\phi_{m}|=1, \quad m=1,2,...,M. \label{PSC}
\end{align}
The problem \textbf{(P2)}
is further decoupled into two sub-problems, one for power allocation at the transmitter, and the other one for RIS's phase shift design. Specifically, the sub-problem for power allocation at the transmitter can be written as
 \begin{align}\label{sub_one}
 	\textbf{(P2.1)} \quad \quad &\max_{\mathbf{E}_k} \quad C_k=\sum_{k=1}^K \log(1+\delta_k),	\\
 	& \quad  s.t. \quad \sum_{k=1}^{K}||\bm{E}_{k}||^{2}\leq P_{T}.
 \end{align}

As discussed before, with sub-precoding matrices $\mathbf{Q}$ and $\mathbf{D}$, the considers multi-user MIMO system can be seen as a virtual MIMO system with distributed receiving antennas, where transmitter has $N$ transmit antennas and receiver has $KL$ receiving antennas. And the above operations open up the possibility of enforcing a simple power allocation policy, in which each individual symbol is allocated a different power. In this work, the optimal water filling policy \cite{Water-filling} for point-to-point MIMO system has been considered, details are shown in the following.

Specifically, the power weighting factors can be expressed
as block-matrices of the global prescaling matrix $\mathbf{E}$ that will be
applied directly to the transmitted symbols, defined as

\begin{equation}
	\begin{aligned} \label{11}
		\mathbf{E}
		=\diag
		\left[
\mathbf{E}_1,\mathbf{E}_2,...,\mathbf{E}_K
		\right],
	\end{aligned}
\end{equation}
with $\mathbf{E}_k \in \mathbb{C}^{K \times K}$, $k=1,2,...,K$ being a diagonal matrix with diagonal elements that represent the power adjustment coefficients for receiver $k$. The power scaling  matrix $\mathbf{E}$ will multiply the data symbol vector before applying the subsequent inter-user OAM mode cancellation matrix and the inter-mode self-interference elimination matrix. By solving the multidimensional problem with the well-known Lagrange
multiplier method, which leads to the following solutions \cite{Water-filling}:
\begin{equation}
	\begin{aligned} \label{decom1}
		E_{k,k}= \frac{P_T}{K}+\frac{1}{K}\sum_{k=1}^{K}\frac{\sigma^2}{\lambda_{k,k}^2}-\frac{\sigma^2}{\lambda_{k,k}^2},
	\end{aligned}
\end{equation}
where $E_{k,k}$ represents the $k$-th diagonal elements of user $k$'s power allocation matrix $\mathbf{E}_k$, and $\lambda_{k,k}$ is the $k$-th singular value of the channel  $\mathbf{\Gamma}_k\mathbf{Q}_k\mathbf{D}_k$.

\section{RIS's Phase Shift Pattern Design for Multi-user OAM Communication Systems } 
 With updated precoder design shown in Section \ref{Section11}, we aim to design the RIS's phase shift pattern for sum rate maximization in this section.
It should be noted that with the proposed three-step precoder design, the inter-mode self-interference and the inter-user OAM mode interference can be completely eliminated, thereby the whole system can be seen as a multiple-input multiple-output (MIMO) communication scenario with $N_T$ transmit antennas and $KN_R$ receiving antennas \cite{Capacity_chara,Achievable_rate}. Thereby the phase shift optimization problem with given precoding scheme can be rewritten as
\begin{align}
	\textbf{(P2.2)} \quad \quad &\max_{\mathbf{\Phi}} \quad 
	\log_{2} \,\det \left(
	\mathbf{I}_{N}+\frac{1}{\sigma^2}\mathbf{\Gamma}\mathbf{R}\mathbf{\Gamma}^H
	\right),\label{sub_two1}	\\
	& \quad \quad \quad \quad |\phi_{m}|=1, \quad m=1,2,...,M. \label{PSC1}
\end{align}
with $\mathbf{R}=\mathbf{W}\mathbf{W}^H$ being the covariance matrix of the transmitted signals and $\mathbf{\Gamma}=\mathbf{B} \mathbf{\Phi}\mathbf{H}\mathbf{F}$ denoting the equivalent channel. The problem \textbf{(P2.2)}
is a non-convex optimization problem since the objective function can be shown to be non-convex over the reflection matrix $\mathbf{\Phi}$, and the unit module constraint on each reflection coefficient in equation \eqref{PSC1} is also non-convex. In the next, we will solve \textbf{(P2.2)} by exploiting its unique structure.
The main idea of solving  \textbf{(P2.2)} is to iteratively solve a series of subproblem, each aiming to optimize one single variable in $\pmb{\phi}=[\phi_{1},\phi_{2},...,\phi_{M}]^T$ with the other $(M-1)$ variables being fixed. To begin with, we first give a more tractable expression for the objective function of \textbf{(P2.2)} in terms of $\phi_m$, $m=1,2,...,M$, since the relationship between $\phi_m$ and the objective function is not explicit. Thus, we aim to rewrite the objective function \eqref{sub_two1} as an explicit function over $a_m$. First, we denote $\mathbf{Z}=\mathbf{H}\mathbf{F}$, and $\mathbf{Z}=[\bm{z}_1, \bm{z}_2,...,\bm{z}_M]^H$, where $z_m^H$ denotes the $m$-th row of the matrix $\mathbf{Z}$. Then the effective MIMO channel from transmitter to all the receivers can be rewritten as 
\begin{equation}
	\begin{aligned} \label{Re_channel}
	\mathbf{\Gamma}=\sum_{m=1}^M \phi_m \bm{b}_m\bm{z}_m^H,
	\end{aligned}
\end{equation}
where $\bm{b}_m \in \mathbb{C}^{KK \times 1}$ represents the $m$-th column of $\mathbf{B}=\left[\mathbf{F}_K^H\mathbf{G}_1,
\mathbf{F}_K^H\mathbf{G}_2,\cdots,\mathbf{F}_K^H\mathbf{G}_K
\right]^T$. Note that from equation \eqref{Re_channel} that the effective channel is in fact a $M$ rank-one matrices $\bm{b}_m\bm{z}_m^H$, each multiplied by a reflection coefficient $\phi_m$, which is a unique structure of RIS-assisted communication channel and implies that $\phi_m$, $m=1,2,...,M$ should be designed to strike an optimal balance between the $M$ matrices for maximizing
the system capacity.

Furthermore, we denote $\mathbf{R}=\mathbf{U}_R\mathbf{\Sigma}_R \mathbf{U}_R^H$ as the eigenvalue decomposition (EVD) of $\mathbf{R}$, where $\mathbf{U}_R \in \mathbb{C}^{N_T \times N_T}$ and $\mathbf{\Sigma}_R \in \mathbb{C}^{N_T \times N_T}$. Besides, since $\mathbf{R}$ is a positive semi-definite matrix, all the diagonal elements in $\mathbf{\Sigma}_R$ are non-negative real numbers. Based on this, we define $\bar{\mathbf{\Gamma}}=\mathbf{\Gamma}\mathbf{U}_R\mathbf{\Sigma}_R^{1/2}$ and $\bar{\mathbf{Z}}=\mathbf{Z}\mathbf{U}_R\mathbf{\Sigma}_R^{1/2}$, therefore, the objective function of \textbf{(P2.2)} can be rewritten as
	\begin{align}
 \label{Re_Objective}
		f \!\!
	&	\triangleq\!\!
	\log_{2} \! \det\! \left(
	\mathbf{I}_{N}+\frac{1}{\sigma^2}\mathbf{\Gamma}\mathbf{R}\mathbf{\Gamma}^H
	\right) \notag \\
&	\!=\!\!
	\log_{2} \!\det \!\left(
	\mathbf{I}_{N}+\frac{1}{\sigma^2}\left(\bar{\mathbf{\Gamma}}\mathbf{U}_R\mathbf{\Sigma}_R^{1/2}\right)\left(\bar{\mathbf{\Gamma}}\mathbf{U}_R\mathbf{\Sigma}_R^{1/2}\right)^H
	\right) \notag \\
&	\!=\!\!\log_{2} \!\det \!\left(
	\mathbf{I}_{N}+\frac{1}{\sigma^2}
	\left(
	\mathbf{B}\mathbf{\Phi}\bar{\mathbf{Z}}
	\right)
		\left(
	\mathbf{B}\mathbf{\Phi}\bar{\mathbf{Z}}
	\right)^H
	\right) \notag \\
&	\!=\!\!\log_{2} \!\det\! \left(\!
	\mathbf{I}_{N}\!+\!\frac{1}{\sigma^2}
	\left(\!
	\sum_{m=1}^M \phi_m \bm{b}_m(\bm{z}_{m}^{'})^H\!\!
	\right)\!\!
	\left(\!
	\sum_{m=1}^M \phi_m \bm{b}_m(\bm{z}_{m}^{'})^H\!\!
	\right)^{\!\!\!\!H}
	\right) \notag \\
&	\!\overset{(a)}{=}\!\!
\log_{2} \!\det\! \left(
\mathbf{I}_{N}+\frac{1}{\sigma^2}\sum_{m=1}^M\bm{b}_m(\bm{z}_{m}^{'})^H\bm{z}_{m}^{'}\bm{b}_m^H \right. \notag  \\ & \left. \quad\quad\quad\quad\quad\quad+\frac{1}{\sigma^2}\sum_{i=1}^M\sum_{j=1,j\neq i}^M \phi_i\phi_j
\bm{b}_m(\bm{z}_{m}^{'})^H\bm{z}_{m}^{'}\bm{b}_m^H
\right),
	\end{align}
where $(a)$ holds due to the fact $|\phi_m|^2=1$, and $(z_m^{'})^H$ denotes the $m$-th row of the matrix $\bar{\mathbf{Z}}$. Note that the new
objective function shown in \eqref{Re_Objective} is an explicit form of the individual reflection coefficients $\phi_m$, $m=1,2,...,M$, which facilitates the following alternating optimization. To proceed further, we rewrite the objective function \eqref{Re_Objective} as the following form with respect to each $\phi_m$:
\begin{equation}
	\begin{aligned} \label{Re_Objective11}
		f_m = &
		\log_2\det
		\left(
		\mathbf{J}_m+\phi_m\mathbf{O}_m+\phi_m^{*}\mathbf{O}_m^H
		\right), \\
		&\quad\quad\quad\quad\quad\quad\quad\quad\quad\quad \forall m=1,2,...,M.
	\end{aligned}
\end{equation}
where 
\begin{equation}
	\begin{aligned} \label{Re_Obj22}
		\mathbf{J}_m=& \,\mathbf{I}_{KK}\\& +\!\! \frac{1}{\sigma^2}\!\!
		\left(\!\!
		\bar{\mathbf{\Gamma}}\!+\!\!\!\!\!\!\sum_{i=1,i\neq m}^M\!\!\!\!\! \phi_m \bm{b}_m(\bm{z}_{m}^{'})^H
		\right)\!\!\!\!
		\left(\!\!
		\bar{\mathbf{\Gamma}}\!+\!\!\!\!\!\sum_{i=1,i\neq m}^M \!\!\!\!\!\!\phi_m \bm{b}_m(\bm{z}_{m}^{'})^H
		\right)^H\\
		&+ \frac{1}{\sigma^2}\bm{b}_m(\bm{z}_{m}^{'})^H\bm{z}_{m}^{'}\bm{b}_m^H, \quad \forall m=1,2,...,M.
	\end{aligned}
\end{equation}
and 
\begin{equation}
	\begin{aligned} \label{Re_Obj33}
		\mathbf{O}_m=\frac{1}{\sigma^2}\bm{b}_m(\bm{z}_{m}^{'})^H
		\left(
		\bar{\mathbf{\Gamma}}^H + \sum_{i=1,i\neq m}^M
\bm{z}_{m}^{'}	\bm{b}_m^H	\phi_i^{*}	\right), \\\quad \forall m=1,2,...,M.
	\end{aligned}
\end{equation}
Therefore, the subproblem  \textbf{(P2.2)} shown in equation \eqref{sub_two1} can be further expressed as 
\begin{align}
	\textbf{(P2.2.1)} \quad \quad &\max_{\phi_m} \quad 
	\log_2\det
	\left(
	\mathbf{J}_m+\phi_m\mathbf{O}_m+\phi_m^{*}\mathbf{O}_m^H
	\right),\label{sub_two11}	\\
	& \quad \quad \quad  |\phi_{m}|=1, \quad m=1,2,...,M. \label{PSC11}
\end{align}
Form \textbf{(P2.2.1)}, we can see that both $\mathbf{J}_m$ and $\mathbf{O}_m$ are independent of $\phi_m$, and the optimal solution of $\phi_m$ is given by \cite{Capacity_chara}, which can be expressed as 
\begin{equation}
	\begin{aligned} \label{Optimal}
		\phi_m^{*}= e^{-j \arg(\epsilon_m)},
	\end{aligned}
\end{equation}
where $\epsilon_m$ is the only non-zero eigenvalue of $\mathbf{J}_m^{-1}\mathbf{O}_m$, since we can seen from equation \eqref{Re_Obj33} that the rank of $\mathbf{O}_m$ is equal to $1$.

\begin{algorithm}[t]
	\caption{- Joint Precoder and Reflection Design for RIS-assisted Multi-user OAM Communication Systems}
	\label{alg:SA}
	\begin{algorithmic}[1]\label{alg:algorithm1}
		\STATE \textbf{Input:} The corresponding channel $\mathbf{G}_k$ from RIS to user $k$, the channel $\mathbf{H}$ from the transmitter to RIS, the transmit power constraint $P_T$ and the noise power $\sigma^2$, the algorithm convergence threshold $\epsilon$, the maximum iteration time $I_{max}$.\\
		\STATE \textbf{Output:} The precoder matrix $\mathbf{W}$ and the phase shift matrix $\mathbf{\Phi}$.
		\STATE \textbf{Initialization}: Randomly generate $\phi_m$ that satisfies the constraint $|\phi_m|=1$, set the algorithm iteration time $i=1$.
		\STATE \textbf{Stage One: Precoder Design}  \\
		\STATE 	 Calculating the inter-user OAM mode interference elimination matrix $\mathbf{Q}$, the inter-mode self-interference cancellation matrix $\mathbf{D}$ and the power allocation matrices $\mathbf{E}$ according to equations \eqref{decom}, \eqref{inter-mode} and \eqref{decom1}, accordingly.\\
		\STATE 		  Generating the overall precoder matrix $\mathbf{W}$ according to equation $\mathbf{W}=\mathbf{Q}\mathbf{D}\mathbf{E}$.\\
		\STATE \textbf{Stage Two: RIS Phase Shift Pattern Optimization}  \\
		
		\FOR{$m=1 \rightarrow M$}
		\STATE Calculating $\mathbf{J}_m$ and $\mathbf{O}_m$ according to equations \eqref{Re_Obj22} and \eqref{Re_Obj33}, respectively,
		\STATE Obtaining the optimal solution of $\phi_m$, $m=1,2, ..., M$ according to equation \eqref{Optimal},
		\STATE Setting $\mathbf{\Phi}= \diag [\phi_1, \phi_2,...,\phi_M]$.
		\ENDFOR
		\STATE The iteration time $i=i+1$.
		\STATE \textbf{Stage Three: Check Convergence}  \\
		\IF{ The objective function in equation \eqref{OBJ} satisfies $C_k^{(i)} -C_k^{(i-1)} \leq \epsilon $, or the iteration time reaches the maximum value $I_{max}$} 
		\STATE Output the precoder matrix $\mathbf{W}$ and the RIS' phase shift pattern matrix $\mathbf{\Phi}$, end algorithm.
		\ELSE
		\STATE go to \textbf{Stage One}.
		\ENDIF 
		
	\end{algorithmic}
\end{algorithm}

The overall algorithm for joint precoder and reflection design for RIS-assisted multi-user OAM communication systems is organized in \textbf{Algorithm 1}. At first, we randomly generate a set of $\phi_m$, $m=1,2,..,M$, that satisfies $|\phi_m|=1$, the algorithm then proceeds by iteratively solving the precoder design sub-problem and the RIS's phase shift pattern design sub-problem until the convergence threshold is reached. Note that in \textbf{Algorithm 1}, we have obtained the optimal solutions of each subproblem. Thereby, the monotonic convergence of \textbf{Algorithm 1} is guaranteed. Besides, since the algorithm yields non-decreasing objective value of \textbf{(P1)} over the iterations, which is also upper-bounded by the finite transmit power $P_T$, thereby \textbf{Algorithm 1} is guaranteed to converge to at least a locally optimal solution. 

The computing complexity of \textbf{Algorithm 1} mainly comes from calculating the sub-precoding matrices $\mathbf{Q}$, $\mathbf{D}$ and $\mathbf{E}$, and generating the matrices $\mathbf{J}_m$ and $\mathbf{O}_m$ for updating $\phi_m$. The computational complexity for calculating the first sub-precoding matrix $\mathbf{Q}$ depends on the SVD operation shown in equation \eqref{decom}, and its computational complexity is $\mathcal{O}\left( K^6\right)$ \cite{Complexity1}. The computational complexity for calculating the second sub-precoding matrix $\mathbf{D}$ depends on the inverse and SVD operations, which has the computational complexity of $\mathcal{O}\left( K^3\right)$ \cite{Complexity1}. And we use water-filling method to obtain the third sub-precoding matrix $\mathbf{E}$ with computational complexity $\mathcal{O}\left( K^2\right)$ \cite{Complexity2}. Combined all the above, the overall complexity for precoder design is $\mathcal{O}\left(K^6+K^3+ K^2\right)$. For updating the phase shift at the RIS, the main computational complexity lies on the inverse operation of $\mathbf{J}_m$ required in equation \eqref{Optimal}, the inverse operation of $\mathbf{J}_m$ has the complexity of $\mathcal{O}\left( N_T^3\right)$. Thereby the final computational complexity of \textbf{Algorithm 1} is $\mathcal{O}\left(N_T^3 +K^6+K^3+ K^2\right)$.

\section{Performance Evaluation}
This section provides numerical results to show the effectiveness of the proposed low-complexity scheme for joint precoder and reflecor design for RIS-assisted multi-user OAM communication systems.
In the simulations, we assume that the transmit UCA has $N_T=20$ antennas, which is located at the origin of the 3D coordinate system.
And there are $K=4$ users, each is equipped with $5$ antennas, thereby the requirement $N_T=KN_R$ for UCA-based multi-user OAM communications can be satisfied \cite{Multi-user,Joint_spatial, Achievable_rate_maximization}. Besides, all the users are randomly distributed in a sphere centered at (10 $\mathrm{m}$, 2 $\mathrm{m}$, 1 $\mathrm{m}$) with the radius of 2 $\mathrm{m}$. The radii of transmit UCA and receive UCAs are set to be $R_t=1$ $\mathrm{m}$ and $R_r= 0.2$ $\mathrm{m}$, respectively.
The RIS is equipped with $M=60$ reflecting elements and it's center point is located at (5 $\mathrm{m}$, 2 $\mathrm{m}$, 1 $\mathrm{m}$). The system is assumed to operate at the carrier frequency of $5$ GHz. Thus, the corresponding wavelength is $\lambda = 0.06 \, \mathrm{m}$. The constant $\beta$ is set to $1$ which means there is no additional antenna gain here \cite{Joint_reflection,Quasi-fractal,Multi-beam}. 
The element spacing at RIS is set to $d_y=d_z=\lambda/2$. 
In this section,
we first investigate the convergence property of the proposed design for joint precoder and RIS phase shift design shown in \textbf{Algorithm 1}. Then we compare the proposed design for multi-user OAM communication systems with other benchmark schemes. Details are shown in the following.


\begin{figure}[t]
	\centering
	\includegraphics[width=0.5\textwidth]{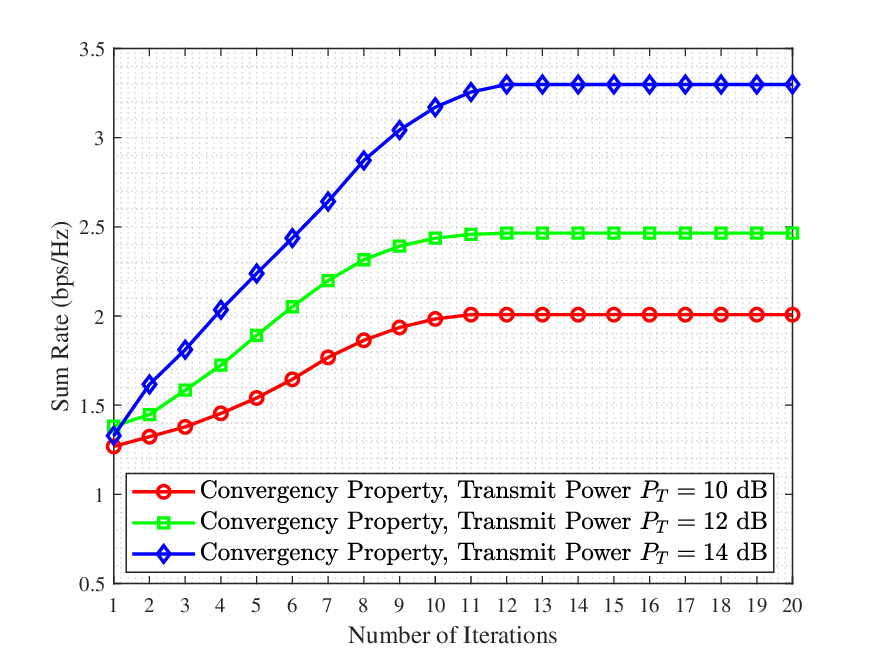}\\
	\caption{Convergence behavior of the joint precoder and reflector design for multi-user OAM communication systems with different transmit powers. The reflecting elements at the RIS is $M=60$, the transmit UCA has $N_T=20$ antennas. And there are $K=4$ receivers, each has $N_R=5$ antennas.}
	\label{Convergence}
\end{figure}

\begin{figure}[t]
	\centering
	\includegraphics[width=0.5\textwidth]{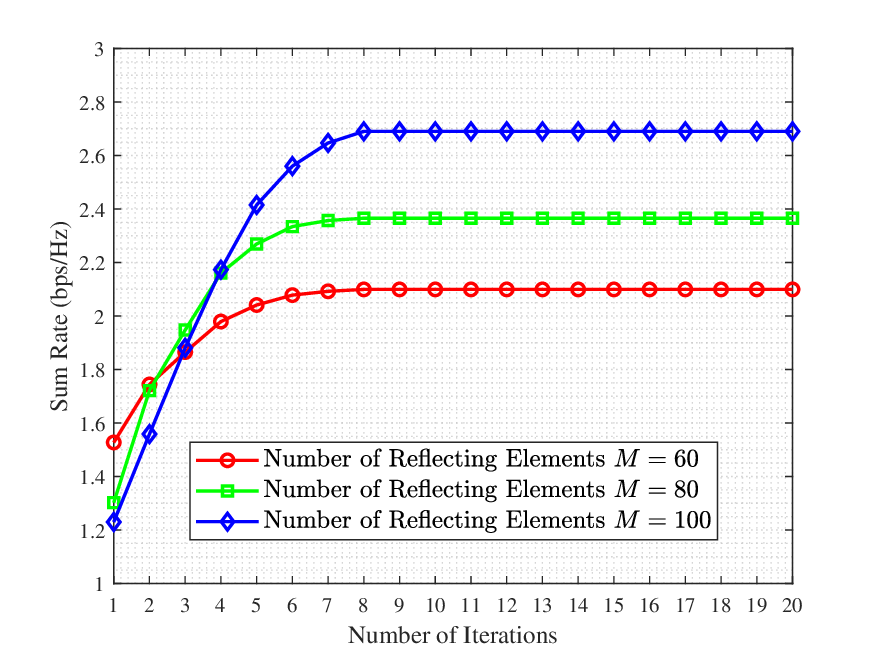}\\
	\caption{Convergence behavior of the joint precoder and reflector design for multi-user OAM communication systems with increasing reflecting elements at the RIS. The transmit power is $P_T=10$ dB, and the transmit UCA has $N_T=20$ antennas, there are $K=4$ receivers, each has $N_R=5$ antennas.}
	\label{M_increasing}
\end{figure}
First, we investigate the convergence behavior and the sum rate performance of the proposed design for multi-user OAM communication systems. In this simulation, we assume that the transmitter has $N_T=20$ antennas. There are $K=4$ receivers, each of them has $N_R=5$ antennas. The number of reflecting elements at the RIS is set to be $M=60$. Fig. \ref{Convergence} shows the convergence behavior of the proposed low-complexity design for multi-user OAM communication systems. It can be observed from Fig. \ref{Convergence} that the proposed algorithm converges as stated and converges fast under different transmit powers.
In Fig. \ref{M_increasing}, we further investigate the impact of the number of RIS's reflecting elements on the sum rate performance. When simulating Fig. \ref{M_increasing}, we set the number of transmit antennas $N_T=20$, and there are $K=4$ users, each of them has $N_R=5$ antennas. The number of RIS's reflecting elements is set to be $M=60, 80, 100$, accordingly. And the transmit power is $P_T=10$ dB.
It can be observed from Fig. \ref{M_increasing} that the sum rate is getting higher with the increase of the reflecting elements. From both Figs. \ref{Convergence} and \ref{M_increasing}, we can see that our proposed design for multi-user OAM communication converges very fast, i.e., the convergence point is achieved in about $10$ times of iteration under different settings, which proves that our proposed design has good stability.


To further verify the superiority of our proposed design of multi-user OAM transmission systems, we compare it with the following benchmark schemes:

\begin{itemize}
		
	\item  Traditional MIMO based transmission scheme with UCA antenna structures (UCA-MIMO) \textcolor{red}{add reference}. All the channels are generated in the same way as described in reference \cite{Design_and}, i.e., location-based channel generation. And three classical precoding designs, i.e., the maximum ratio transmission (MRT), zero-forcing (ZF), and minimum mean-squared error (MMSE) precoding schemes are used at the transmitter side. And we still adopting the same phase shift pattern optimization method for RIS as proposed in this work. 
	
	
\end{itemize}

\begin{figure}[t]
	\centering
	\includegraphics[width=0.5\textwidth]{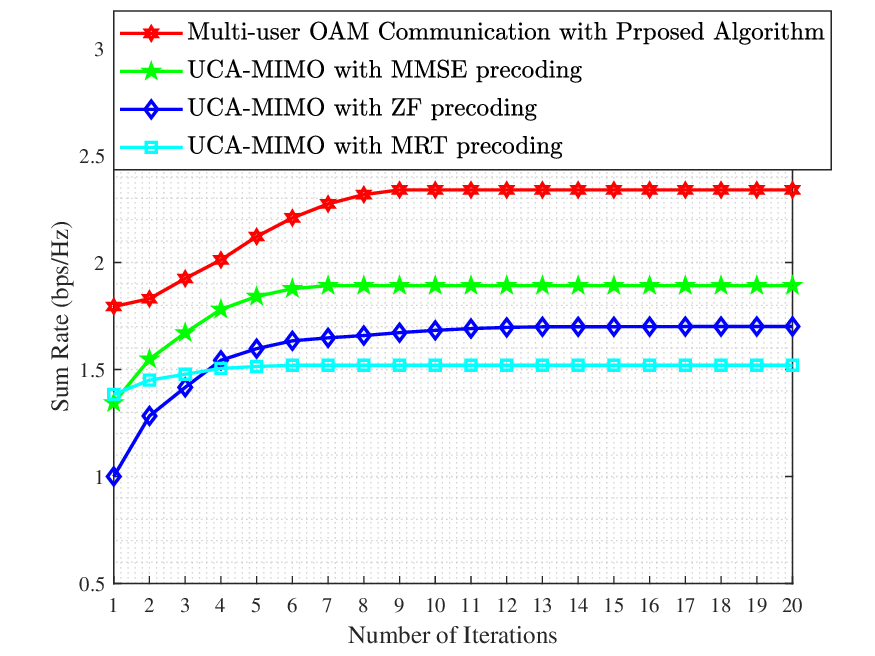}\\
	\caption{ Comparison between the proposed design, traditional MIMO with MRT/ZF/MMSE precoding. The reflecting elements at the RIS is set to be $M=100$. The transmit UCA has $N_T=20$ antennas, there are $K=4$ receivers, each has $N_R=5$ antennas. The transmit power is set to be $P_T=10$ dB.} 
	\label{comparison1}
\end{figure}

 Fig. \ref{comparison1} shows the comparisons between the proposed design and traditional MIMO with MRT/ZF/MMSE precoding, accordingly. 
 When doing the simulation, we set the number of reflecting elements at the RIS to be $M=120$. The transmit UCA has $N_T=20$ antennas, there are $K=4$ receivers, and each has $N_R=5$ antennas.
 It can be seen from Fig. \ref{comparison1} that our proposed design for multi-user OAM transmissions can achieve better sum rate performance since the orthogonality between different modes for different users bring significant performance gains due to the well mitigated inter-user OAM mode interference and inter-mode self-interference. For the traditional UCA-MIMO transmission schemes, we can see from the figure that MMSE scheme has better sum rate performance than ZF and MRT schemes, since MMSE takes both the effects of interference and noise into the considerations. In contrast, the MRT scheme has worse sum rate performance compared with MMSE and ZF, since MRT only considers the channel attenuation of the transmitted signals, which leading to unsatisfactory sum rate performance in multi-user wireless communication systems.

 \begin{figure}[t]
 	\centering
 	\includegraphics[width=0.5\textwidth]{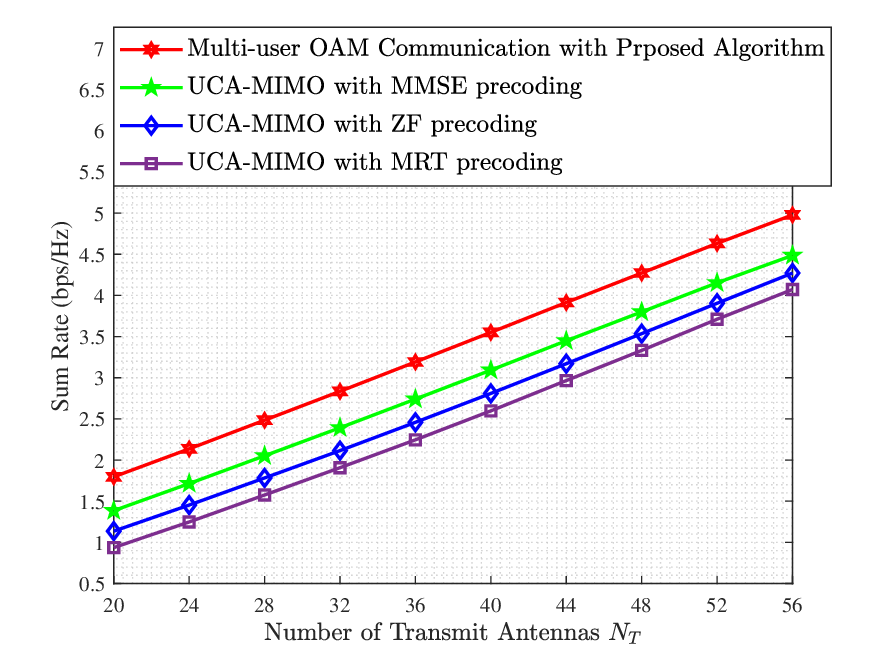}\\
 	\caption{Comparison between the proposed design, traditional MIMO with MRT/ZF/MMSE precoding when the number of transmit antenna increasing. 
 		The reflecting elements at the RIS is set to be $M=80$. There are $K=4$ users. And the transmit power is $P_T=10$ dB.}
 	\label{Comparison2_N_Increasing}
 \end{figure}

 We further compare the sum rate performance of the proposed design and traditional UCA-MIMO with MRT/ZF/MMSE precoding when the number of transmit antenna increasing. In this simulation, the transmit power is set to be $P_T=10$ dB. There are $K=4$ users, each user has $N_R=\frac{N_T}{K}$ receiving antennas, since we need to guarantee that the different OAM modes can be successfully decoded at the receiver side \cite{Multi-user,Joint_spatial}.
  It can be seen from Fig. \ref{Comparison2_N_Increasing} that with the increase of transmit antennas, the sum rate performance increases in all transmission schemes. Among all the schemes, our proposed scheme has better sum rate performance due to the successful inter-user OAM mode and inter-mode self-interference eliminations at the transmitter side, which further verify the superiority of the proposed transmission scheme.
  
     \begin{figure}[t]
  	\centering
  	\includegraphics[width=0.5\textwidth]{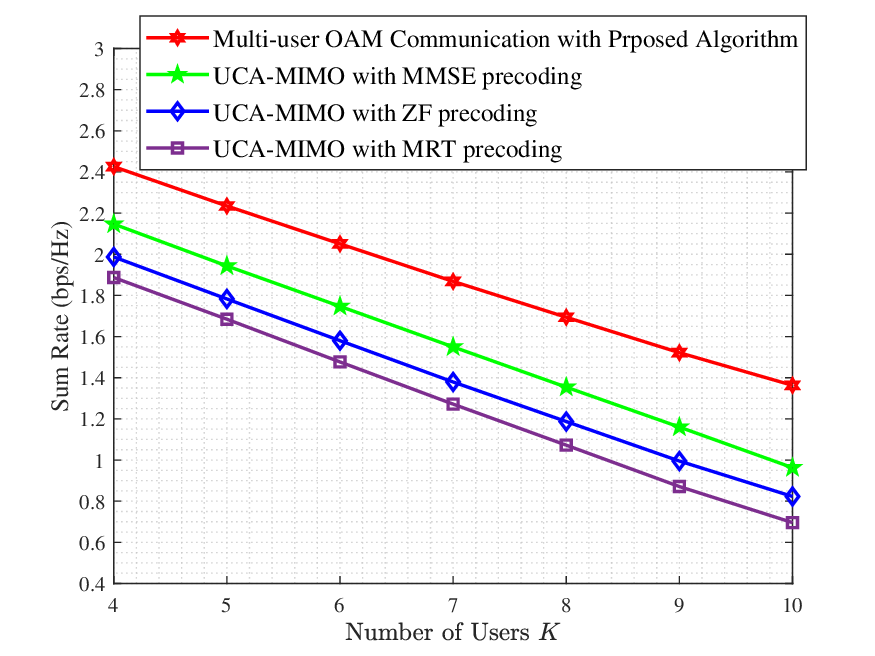}\\
  	\caption{Comparison between the proposed design and traditional MIMO with MRT/ZF/MMSE precoding when the number of users changes. The reflecting elements at the RIS is set to be $M=80$. The transmit power is set to be $P_T=10$ dB.}
  	\label{Comparison_K_Increasing}
  \end{figure}
  Fig. \ref{Comparison_K_Increasing} shows the sum rate performances under different numbers of users. In this simulation, the number of reflecting elements at the RIS is set to be $M=60$ except for the non-RIS-assisted multi-user OAM systems. The transmit power at the transmitter is $P_T=10$ dB. It should be noted that to guarantee different OAM modes can be successfully decoded at the receiver side, $N_T=KN_R$ should be satisfied \cite{Multi-user,Joint_spatial}. Here, we assume that each user has $N_R=5$ antennas. It can be seen from Fig. \ref{Comparison_K_Increasing} that with the increasing number of receivers, the sum rate performance decreases. For our proposed multi-user OAM communication schemes, the performance degradation mainly comes from the reduction of the beamforming gain provided by the RIS, because the RIS needs to balance the performance among all the receivers, thereby the increase of the total receivers will decrease RIS's overall beamforming gains. For traditional UCA-MIMO communication schemes, i.e., UCA- MIMO with MRT/ZF/MMSE precoding, the degradation of the sum rate performance comes from two parts: one is the overall reduction of the beamforming gain provided by the RIS caused by the large number of receivers, the other one is the more severe inter-user interference. Due to the above reasons, we can observe from Fig. \ref{Comparison_K_Increasing} that the performance gap between the proposed multi-user OAM schemes and traditional MIMO schemes, i.e., the performance gap between the proposed scheme (red line with hexagram marks) and traditional MIMO with MMSE precoding (green line with pentagram marks), becomes larger when the number of receivers increases.

    \begin{figure}[t]
	\centering
	\includegraphics[width=0.5\textwidth]{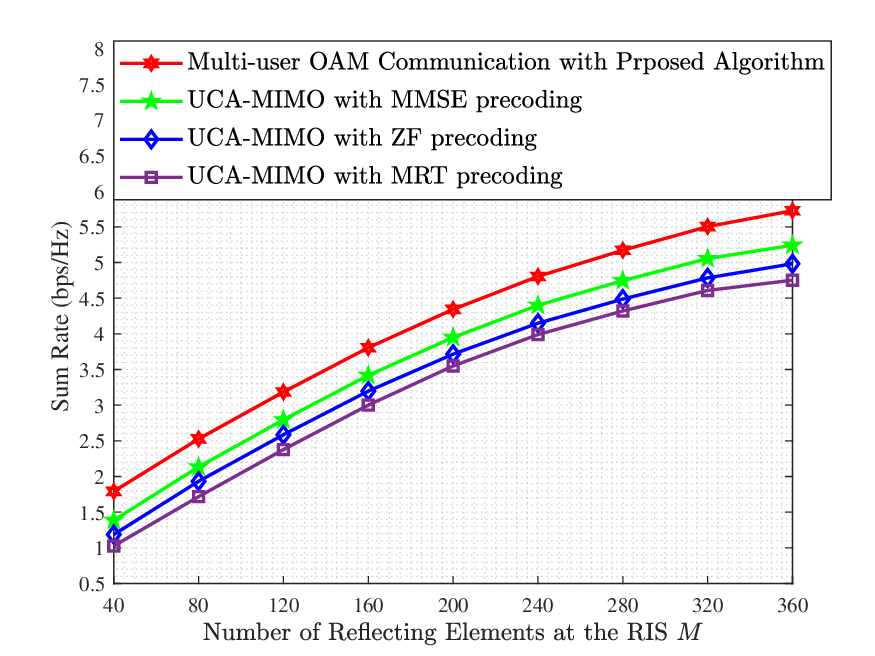}\\
	\caption{Comparison between the proposed design and traditional MIMO with MRT/ZF/MMSE precoding with different numbers of reflecting elements at the RIS. The transmit UCA has $N=20$ antennas, there are $K=4$ receivers, each has $L=5$ antennas. The transmit power is set to be $P_T=10$ dB.}
	\label{comparison3_M_Increasing}
\end{figure}

In Fig. \ref{comparison3_M_Increasing}, we investigate the sum rate performance with the number of RIS's reflecting element $M$ increasing. For this simulation, the transmitter has $N_T=20$ transmit antennas, and there are $K=4$ receivers, each receiver has $N_R=5$ receiving antennas. From Fig. \ref{comparison3_M_Increasing}, we can see that with the number of reflecting elements increasing, all the schemes except non-RIS-assisted multi-user OAM transmission can achieve higher sum rate performance. Among all the schemes, our proposed multi-user OAM transmission design can always generate better sum rate performance due to the will-eliminated inter-user and inter-mode interference. 

	\section{Conclusions}
	In this paper, we consider a RIS-assisted multi-user OAM communication systems by taking the phase turbulence into consideration.
 We formulate a sum rate maximization problem for multi-user OAM systems through the precoder design and the RIS's phase shift optimization.
 For the precoder matrix design, we propose a three-layer scheme, which includes inter-user interference cancellation, inter-mode interference elimination and the power allocation. For inter-user and inter-mode interference management, we directly give the specific  expressions by analyzing the characteristics of the overall channel, which can significantly reduce the complexity for system design. And finally the power allocation is achieved trough traditional water-filling methods. 
 With the proposed three-layer precoder design, we are able to transform the multi-user OAM transmission system into a virtual MIMO system with distributed receiving antennas. By further analyzing the overall channel model, we are able to obtain the closed-form expressions for updating the RIS's phase shift design, which can avoid the high optimization complexity caused by the large number of reflecting elements at the RIS.
 We also compare the proposed multi-user OAM communication scheme with traditional MIMO transmission schemes, numerical results have shown that our proposed multi-user OAM transmission scheme can achieve better sum rate performance due to the well-mitigated inter-user and inter-mode interference.

	%
	
  \begin{spacing}{1.232}
	\bibliographystyle{plain}
	\bibliography{myRef}
    \end{spacing}

\end{document}